\documentclass[conference]{IEEEtran}

\usepackage{amsmath,mathtools}
\usepackage{amssymb}
\usepackage{pifont}
\usepackage{fancyhdr}
\usepackage{graphicx}
\usepackage{amsmath}
\usepackage{algorithm}
\usepackage{subfig}

\usepackage{algpseudocode}
\usepackage{boxedminipage}
\usepackage{fancyhdr}
\usepackage{multirow}

\usepackage{array}
\usepackage{textcomp}
\usepackage{stfloats}
\usepackage{csquotes} 
\usepackage{url}
\usepackage{verbatim}
\usepackage{cite}
\usepackage{eqnarray}
\def\BibTeX{{\rm B\kern-.05em{\sc i\kern-.025em b}\kern-.08em
    T\kern-.1667em\lower.7ex\hbox{E}\kern-.125emX}}
\usepackage{balance}
\usepackage{url}

\begin{document}

\title{A Modular DTaaS Architecture for Predictive Slice Management in 6G Systems}

\author{
    \IEEEauthorblockN{Tuğçe BILEN\IEEEauthorrefmark{1} and Mehmet Ozdem\IEEEauthorrefmark{2}}
    
    \IEEEauthorblockA{\IEEEauthorrefmark{1}Department of Artificial Intelligence and Data Engineering\\Faculty of Computer and Informatics\\
    Istanbul Technical University, Istanbul, Turkey\\}
    \IEEEauthorblockA{\IEEEauthorrefmark{2}Turk Telekom, Ankara, Turkey \\
    Email: bilent@itu.edu.tr, mehmet.ozdem@turktelekom.com.tr}
}

% make the title area
\maketitle

\begin{abstract}
The sixth generation (6G) of wireless networks will require fundamentally new orchestration paradigms to meet stringent requirements for ultra-low latency, high reliability, and pervasive intelligence. Network slicing emerges as a key enabler to support diverse services with customized quality-of-service (QoS) guarantees. However, dynamic and fine-grained slice management poses significant challenges in terms of real-time provisioning, SLA assurance, and cross-layer observability. In this paper, we propose a novel Digital Twin as a Service (DTaaS) framework that embeds per-slice digital twins (SDTs) into the orchestration loop. Each SDT maintains a synchronized, real-time representation of its slice, leveraging multi-domain telemetry and deep sequential models to predict traffic evolution and SLA risks. The framework introduces modular intelligence layers, programmable interfaces, and edge-embedded decision-making to enable proactive provisioning, adaptive scaling, and closed-loop SLA assurance. Mathematical formulations for fidelity measurement, predictive control, and optimization objectives are provided to ensure rigor and transparency. Evaluation results demonstrate that DTaaS significantly improves SLA compliance ratio, reduces resource over-provisioning, and lowers average SLA violation probability, offering a scalable and reliable orchestration approach for 6G networks.
\end{abstract}


\section{Introduction}
The sixth generation (6G) of wireless networks is poised to revolutionize communication systems by offering extreme performance capabilities, including sub-millisecond latency, ultra-high reliability, and pervasive intelligence \cite{10215892}. As 6G evolves to support advanced applications such as aeronautical communications, extended reality, and industrial automation, it demands a fundamental rethinking of network control and service orchestration paradigms \cite{9520342}. Unlike previous generations, 6G is expected to be inherently AI-native, context-aware, and zero-touch, where real-time adaptability is a mandatory design principle rather than an optimization goal.

Among the key architectural enablers of 6G, network slicing has emerged as a crucial mechanism for supporting service diversity and isolation. A network slice is a logical end-to-end construct that allocates radio, computing, and storage resources to meet the specific quality-of-service (QoS) requirements of diverse service classes, including enhanced Mobile Broadband (eMBB), ultra-Reliable Low-Latency Communication (URLLC), and massive Machine-Type Communication (mMTC). This logical partitioning allows network operators to provide customized performance guarantees, maximize resource efficiency, and ensure that each service category receives the necessary quality \cite{BILEN201824}.

However, as network slicing becomes increasingly dynamic and fine-grained in 6G, its real-time provisioning and lifecycle management introduce significant challenges. First, most existing slicing solutions operate reactively, allocating or scaling resources only after congestion or degradation has occurred. This delayed response often leads to service-level agreement (SLA) violations or degraded quality of experience (QoE), particularly in environments with bursty or rapidly changing traffic. Second, conventional orchestrators lack a continuously evolving digital representation of slice states, making it difficult to anticipate future conditions or preemptively adapt network behavior. Third, limited observability and weak semantic integration across heterogeneous telemetry sources further hinder closed-loop automation and fine-grained decision-making.

To overcome these limitations, we propose an innovative architectural framework built upon the concept of Digital Twin as a Service (DTaaS). Our approach instantiates and maintains real-time digital representations, referred to as Slice Digital Twins (SDTs), for each active network slice and integrates them as first-class entities within the orchestration loop. By embedding predictive intelligence within SDTs, the network gains the ability to anticipate load dynamics, proactively reserve resources, and enforce SLA guarantees before violations occur. Unlike traditional digital twin implementations, which are typically focused on physical assets or isolated IoT environments, our framework emphasizes end-to-end network abstraction and service-level orchestration. It introduces modular, API-interoperable twin services that are lightweight enough for deployment at the edge and capable of consuming and processing cross-layer telemetry in real time. As a result, the proposed approach enables proactive slice management and significantly improves the reliability, efficiency, and adaptability of 6G networks. The key contributions of this work are summarized as follows:

\begin{itemize}
    \item We propose a modular DTaaS framework that embeds digital twins into the orchestration loop of 6G network slicing for real-time observability and adaptive control.
    \item We define per-slice digital twins that leverage real-time telemetry and deep sequential models to predict SLA risks and forecast load dynamics.
    \item We design a scalable twin orchestration layer with semantic APIs to support closed-loop control, SLA tracking, and efficient lifecycle management.
    \item We implement a twin-driven provisioning agent that enables proactive scaling and resource reallocation based on predicted slice demands, thereby enhancing SLA adherence.
\end{itemize}

The remainder of the paper is organized as follows. Section II reviews related work. Section III presents the proposed DTaaS architecture and slice management model. Section IV evaluates the performance of the proposed system, and Section V concludes the paper.

\section{Related Work}

The evolution of 6G networks has spurred extensive research into network slicing, digital twins, and AI-driven orchestration to meet stringent performance requirements. This section reviews key works from IEEE Xplore, highlighting their contributions, limitations, and how our proposed DTaaS framework advances the state of the art in predictive slice management.

Network slicing is a foundational technology for service differentiation in 5G and 6G. In \cite{ref1}, a reinforcement learning (RL) approach is proposed for network slicing in 5G networks, enabling dynamic resource allocation across diverse services. While it enhances efficiency, the method is reactive and may incur SLA violations in highly dynamic 6G environments. Similarly, \cite{ref2} introduces a hybrid deep learning model for enhanced network slicing in 6G, focusing on security, latency reduction, and throughput improvement. However, it lacks integration with real-time digital representations for proactive forecasting. Our DTaaS framework overcomes these by embedding per-slice digital twins (SDTs) with predictive models for proactive provisioning and SLA assurance.

Digital twins (DTs) offer promising capabilities for network optimization. In \cite{ref3}, a digital-twin-driven approach is presented for end-to-end network slicing toward 6G, integrating AI to enable slicing orchestration. The work demonstrates improved management but does not emphasize edge deployment or multi-domain telemetry for fine-grained predictions. Likewise, \cite{ref4} proposes a modeling and deployment framework for digital twins in 5G networks, creating virtual replicas for monitoring and optimization. Although applicable to 6G, it focuses on physical layer aspects without addressing service-level slicing dynamics. Our DTaaS extends DTs to a modular, edge-embedded service, incorporating cross-layer telemetry for slice-specific predictive orchestration.

\begin{table}[ht]
\centering
\caption{Summary of related work}
\label{tab:related}
\begin{tabular}{|p{0.5cm}|p{3.5cm}|p{3.5cm}|}
\hline
\textbf{Ref.} & \textbf{Main Contribution} & \textbf{Limitation / Gap} \\
\hline
\cite{ref1} & RL-based dynamic resource allocation for slicing & Reactive, may cause SLA violations in dynamic 6G \\
\hline
\cite{ref2} & Hybrid DL for slicing with security and latency focus & No integration with real-time digital twins \\
\hline
\cite{ref3} & DT-driven orchestration for end-to-end slicing & Lacks edge deployment and multi-domain telemetry \\
\hline
\cite{ref4} & DT modeling for 5G virtual replicas & Focus on physical layer, not slice-level orchestration \\
\hline
\cite{ref5} & Deep RL for online slice resource allocation & Centralized, unsuitable for URLLC latency \\
\hline
\cite{ref6} & Deep RL for flexible beyond-5G slicing & Static training, lacks continuous feedback \\
\hline
\cite{ref7} & Federated learning for slicing in vehicular networks & No synchronized twin representation for full state tracking \\
\hline
\cite{ref8} & Zero-touch orchestration for large-scale slices & Focus on policies, not predictive SLA management \\
\hline
\end{tabular}
\end{table}

Predictive orchestration is essential for adaptive 6G networks. In \cite{ref5}, deep reinforcement learning is applied for online resource allocation in network slicing, achieving efficient adaptation to varying demands. Its centralized nature, however, introduces latency issues unsuitable for URLLC slices. In \cite{ref6}, deep RL is used for network slicing in beyond-5G and 6G systems to flexibly accommodate wireless services. While scalable, it relies on static training without continuous feedback loops for SLA risk prediction. Our framework deploys lightweight SDTs at the edge, merging sequential models with closed-loop control to reduce latency and bolster SLA compliance.

AI-native paradigms for 6G orchestration have also gained attention. In \cite{ref7}, federated learning is employed for network slicing in 6G systems tailored to autonomous vehicles, improving connectivity through distributed training. The approach enhances privacy but lacks synchronized twin representations for holistic state tracking. Similarly, \cite{ref8} proposes a framework for zero-touch management and orchestration of massive network slice deployments in 6G, supporting automated lifecycle handling. It prioritizes policy enforcement over predictive resource management. Our DTaaS distinguishes itself by integrating SDTs into the orchestration loop, facilitating proactive scaling and reconfiguration via real-time telemetry and analytics.

As summarized in Table \ref{tab:related}, prior works have advanced slicing, digital twins, and AI-driven orchestration, yet they mostly remain reactive, centralized, or limited to partial state monitoring. Our DTaaS framework differs by embedding per-SDTs into the orchestration loop, combining multi-domain telemetry, predictive models, and edge intelligence. This enables proactive resource adaptation and closed-loop SLA assurance, offering a scalable and SLA-centric orchestration paradigm that overcomes key limitations of existing solutions. Moreover, the modular design ensures that each slice can be independently monitored and optimized without cross-slice interference. The integration of sequential prediction models further enhances adaptability by anticipating traffic surges and mobility patterns before they occur. Finally, by deploying lightweight SDTs at the edge, our approach minimizes latency and strengthens reliability, which are critical for demanding 6G services such as URLLC and industrial automation.

   \begin{figure*}[h]
    \centering
    \includegraphics[width=0.55\linewidth]{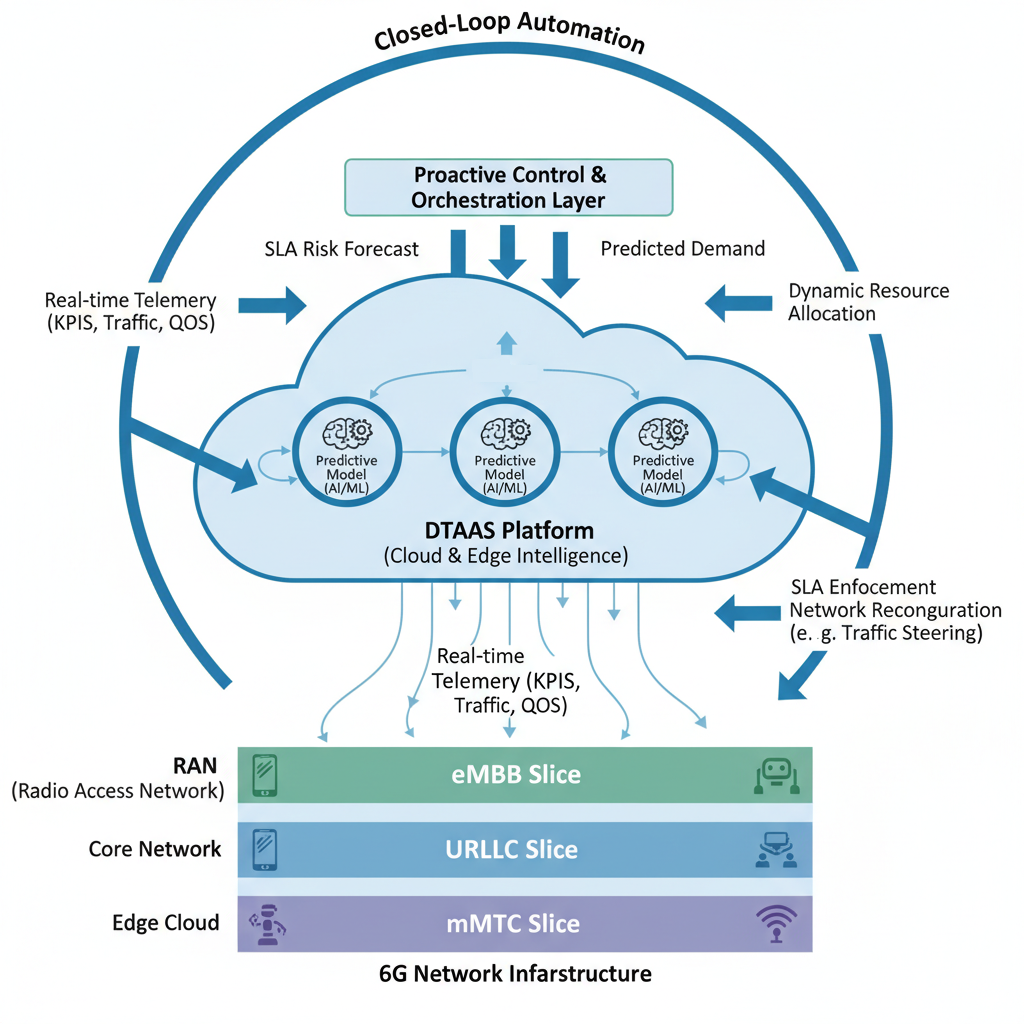}
    \caption{The proposed modular DTaaS architecture.}
    \label{arc}
    \end{figure*}

\section{The Proposed DTaaS Architecture}
\subsection{Design Aspects}
As 6G networks evolve toward ultra-dense, AI-driven, and service-centric infrastructures, digital twins must evolve from static representations into intelligent, active entities directly embedded within the control loop \cite{10078095}. To achieve this, the proposed DTaaS framework introduces modular and predictive twin services designed to enhance slice observability, forecasting capability, and orchestration efficiency. The main design principles are as follows:

\begin{itemize}
    \item {Slice-Centric Twin Mapping:} Each SDT instance is uniquely associated with a specific network slice (eMBB, URLLC, or mMTC), enabling strict isolation, per-slice policy customization, and slice-specific forecasting.
    \item {SLA-Aware Predictive Control:} SDTs integrate sequential inference models that predict traffic evolution and SLA risks, enabling proactive resource adaptation and preventing performance degradation before it occurs.
    \item {Edge-Embedded Intelligence:} Inference and decision-making modules are deployed at edge nodes to minimize control latency, support localized optimization, and reduce reliance on centralized cloud resources.
    \item {Closed-Loop Integration:} SDT outputs directly trigger programmable network actions such as slice scaling, RAN parameter adjustment, or traffic steering, ensuring automated SLA enforcement under dynamic conditions.
    \item {Programmable and Secure Interfaces:} SDT insights are exposed through authenticated APIs, enabling orchestrators and slice tenants to configure policies, monitor performance, and enforce objectives with strict cross-slice isolation.
\end{itemize}

\subsection{System Overview}
The DTaaS architecture is composed of three functional layers that collectively enable end-to-end slice lifecycle management:

\begin{itemize}
    \item {Telemetry and Data Collection Layer:} This layer aggregates multi-domain telemetry, including radio measurements, transport-layer characteristics, and service-level KPIs. The collected data is used to construct and continuously update the Slice Digital Twins.
    \item {Twin Intelligence Layer:} At the core of the framework, this layer hosts per-slice SDTs, each maintaining a synchronized digital representation of slice states and embedding predictive models for SLA risk forecasting. It supports modular plug-ins for traffic load prediction, anomaly detection, and mobility estimation.
    \item {Orchestration and Control Layer:} This layer leverages SDT outputs to make adaptive network decisions. It implements closed-loop control by translating predictions into reconfiguration commands, including dynamic resource scaling and RAN reconfiguration. Secure APIs enable integration with external orchestrators and slice owners.
\end{itemize}

\section{DTaaS-Based Slice Management Model}
As illustrated in Fig. \ref{arc}, the above-explained layers form a continuous feedback cycle linking real-time telemetry, predictive intelligence, and programmable control, enabling proactive slice management and significantly reducing SLA violations. The details of the proposed model is explained in the upcoming parts.

\subsection{Slice Lifecycle with Digital Twins}
In the proposed framework, the lifecycle of each slice is tightly coupled with its corresponding SDT. Once a slice $s_k \in \mathcal{S}$ is instantiated, an SDT is automatically deployed as a containerized service. The slice state is represented by a telemetry vector as given in Eq.~\ref{e1}.

\begin{equation} \label{e1}
    m_k(t) = [\lambda_k(t), \rho_k(t), \gamma_k(t), \eta_k(t)]
\end{equation}

In Eq.~\ref{e1}, $\lambda_k(t)$ denotes the traffic arrival rate, $\rho_k(t)$ is the resource utilization ratio, $\gamma_k(t)$ represents radio channel quality, and $\eta_k(t)$ corresponds to the observed SLA satisfaction ratio at time $t$. The digital twin replica of this state, $\hat{m}_k(t)$, is maintained within the DTaaS system and synchronized continuously as described in Section~III. The \textbf{synchronization fidelity} score is defined as given in Eq.~\ref{e2}.

\begin{equation}\label{e2}
    F_k(t) = \| m_k(t) - \hat{m}_k(t) \|_2
\end{equation}

The orchestration objective is to minimize the aggregate mismatch across all slices as given in Eq.~\ref{e3}.

\begin{equation} \label{e3}
     \min \sum_{k=1}^{K} F_k(t)
\end{equation}

\subsection{Twin-Driven Provisioning and Scaling}
Each SDT integrates a predictive model $\Psi_k(\cdot)$ that forecasts slice demand $h$ steps into the future as $\hat{\lambda}_k(t+h) = \Psi_k(m_k(t), m_k(t-1), \dots)$. If $\hat{\lambda}_k(t+h)$ is predicted to exceed a predefined safety threshold $\Lambda_k^{\text{safe}}$, a proactive provisioning action is triggered. The decision variable for the additional resources allocated to slice $s_k$ is denoted by $\Delta r_k(t)$. The goal is to ensure SLA compliance while minimizing over-provisioning, as formulated in Eq.~\ref{e4}.

\begin{equation} \label{e4}
    \min_{\Delta r_k(t)} \; \alpha \cdot \text{OverAlloc}(\Delta r_k(t)) + \beta \cdot \text{SLA\_Risk}(s_k,t)
\end{equation}

In Eq.~\ref{e4}, $\alpha$ and $\beta$ are weighting factors that reflect operator priorities between resource efficiency and SLA assurance. Also, we explain other components as follows:

\begin{itemize}
    \item $\text{OverAlloc}(\Delta r_k(t)) = \max(0, \Delta r_k(t) - r_k^{\text{needed}}(t))$, where $r_k^{\text{needed}}(t)$ is the estimated minimum resource requirement for slice $k$ at time $t$ derived from $\hat{\lambda}_k(t)$.
    \item $\text{SLA\_Risk}(s_k,t) = \mathbb{P}[\eta_k(t+h) < \theta_k]$ is the probability of SLA violation predicted over the horizon $h$, where $\theta_k$ is the SLA satisfaction threshold for slice $k$.
\end{itemize}

\subsection{Adaptive Reconfiguration and Reallocation}
When localized anomalies or traffic surges are predicted, the SDT proposes reconfiguration actions such as traffic steering or handover parameter adjustments. Let $\pi_k(t)$ denote the configuration policy applied to slice $s_k$. The control objective is to minimize the probability of SLA violations over a prediction horizon $H$, as defined in Eq.~\ref{e5}.

\begin{equation} \label{e5}
    \min_{\pi_k(t)} \sum_{h=1}^{H} \mathbb{P}\big[\eta_k(t+h) < \theta_k\big]
\end{equation}

\subsection{Closed-Loop SLA Assurance}
The DTaaS system operates in a closed-loop manner. After a control decision $\pi_k(t)$ or a provisioning update $\Delta r_k(t)$ is executed, telemetry feedback is collected and compared against SDT predictions. The {prediction error} is defined as given in Eq.~\ref{e6}.

\begin{equation} \label{e6}
    E_k(t) = \| \hat{m}_k(t) - m_k(t) \|_2
\end{equation}

This error is used to refine both the synchronization process and the embedded prediction models $\Psi_k(\cdot)$. This continuous feedback loop enhances the adaptability and robustness of the DTaaS framework.

By embedding predictive models and optimization objectives directly into the orchestration process, the proposed DTaaS model transforms network slicing into a proactive, SLA-centric management paradigm. Provisioning, scaling, and reconfiguration decisions become predictive, mathematically grounded, and continuously validated through closed-loop feedback.

\section{Performance Evaluation}
\subsection{Simulation Environment}
To evaluate the proposed DTaaS framework, a simulation environment was developed in Python using the SimPy discrete-event simulation library \cite{xx9617810}. The testbed emulates a 6G slicing scenario with heterogeneous services, including eMBB, URLLC, and mMTC. Each service category is instantiated as an independent network slice with unique traffic characteristics, latency constraints, and reliability targets.  

The simulated infrastructure comprises three hierarchical domains: edge, transport, and core, interconnected through programmable interfaces that mirror the telemetry and orchestration layers described in Section~III. Slice telemetry, including traffic arrival rate, resource utilization, and observed SLA satisfaction, is generated at one-second intervals. Each SDT operates as a containerized agent that receives telemetry input, performs prediction, and triggers scaling actions through the orchestration layer.

\begin{table}[ht]
\centering
\caption{Simulation Parameters}
\label{tab:params}
\begin{tabular}{|p{3cm}|p{4.5cm}|}
\hline
\textbf{Parameter} & \textbf{Value / Description} \\
\hline
Simulation time & 5000 time slots \\
\hline
Prediction horizon ($h$) & 5 time steps \\
\hline
Traffic model & Poisson arrivals with burst factor = 1.5 \\
\hline
SLA latency thresholds & eMBB: 20~ms, URLLC: 5~ms, mMTC: 50~ms \\
\hline
Edge node capacity & 100 resource units \\
\hline
Twin update interval & 1~s \\
\hline
Seq2Seq hidden units & 64 neurons, learning rate 0.001 \\
\hline
Weighting factors ($\alpha$, $\beta$) & (0.4, 0.6) \\
\hline
\end{tabular}
\end{table}

The DTaaS framework integrates a sequence-to-sequence (Seq2Seq) LSTM model for predictive control. Each SDT is configured to forecast slice demand over a horizon of $h=5$ time steps. To ensure reproducibility, all simulations were executed on a workstation equipped with an AMD Ryzen 9 7950X CPU, 64~GB RAM, and Ubuntu~24.04. The simulation runs for 5000~time slots per scenario, with results averaged over 10~independent repetitions to mitigate random variation. Table~\ref{tab:params} summarizes the main simulation parameters.

\subsubsection{Baseline Methods}
To demonstrate the benefits of the proposed DTaaS framework, two representative baseline approaches were implemented for comparison. These methods were selected because they represent the most common paradigms in recent literature and can be feasibly simulated without full-scale network implementations.

\begin{itemize}
    \item Reactive Slicing Orchestrator (RSO): This baseline represents traditional slicing mechanisms where orchestration decisions are triggered only after SLA degradation is detected. Resource scaling is performed reactively using threshold-based rules. Although simple and widely used, this approach often suffers from delayed adaptation and transient SLA violations.
    \item Centralized Deep Reinforcement Learning (C-DRL): Inspired by recent studies such as \cite{ref5,ref6}, this baseline applies a centralized deep reinforcement learning model trained to allocate resources among active slices. It learns optimal allocation policies over time but lacks real-time predictive awareness and introduces control latency due to centralized inference. Full RL implementations in live networks are complex; therefore, this baseline simulates the logical behavior of centralized agents rather than training on live data, enabling fair comparison with DTaaS.
\end{itemize}

These two baselines collectively capture the dominant directions in the literature (reactive rule-based control and AI-assisted centralized learning) while remaining computationally tractable for simulation. Comparing against these methods allows assessing DTaaS’s key advantages in prediction-driven, distributed, and closed-loop orchestration.

 \begin{figure*}[h]	
	\centering		
	\subfloat[]{%
		\includegraphics[width=0.25\textwidth]{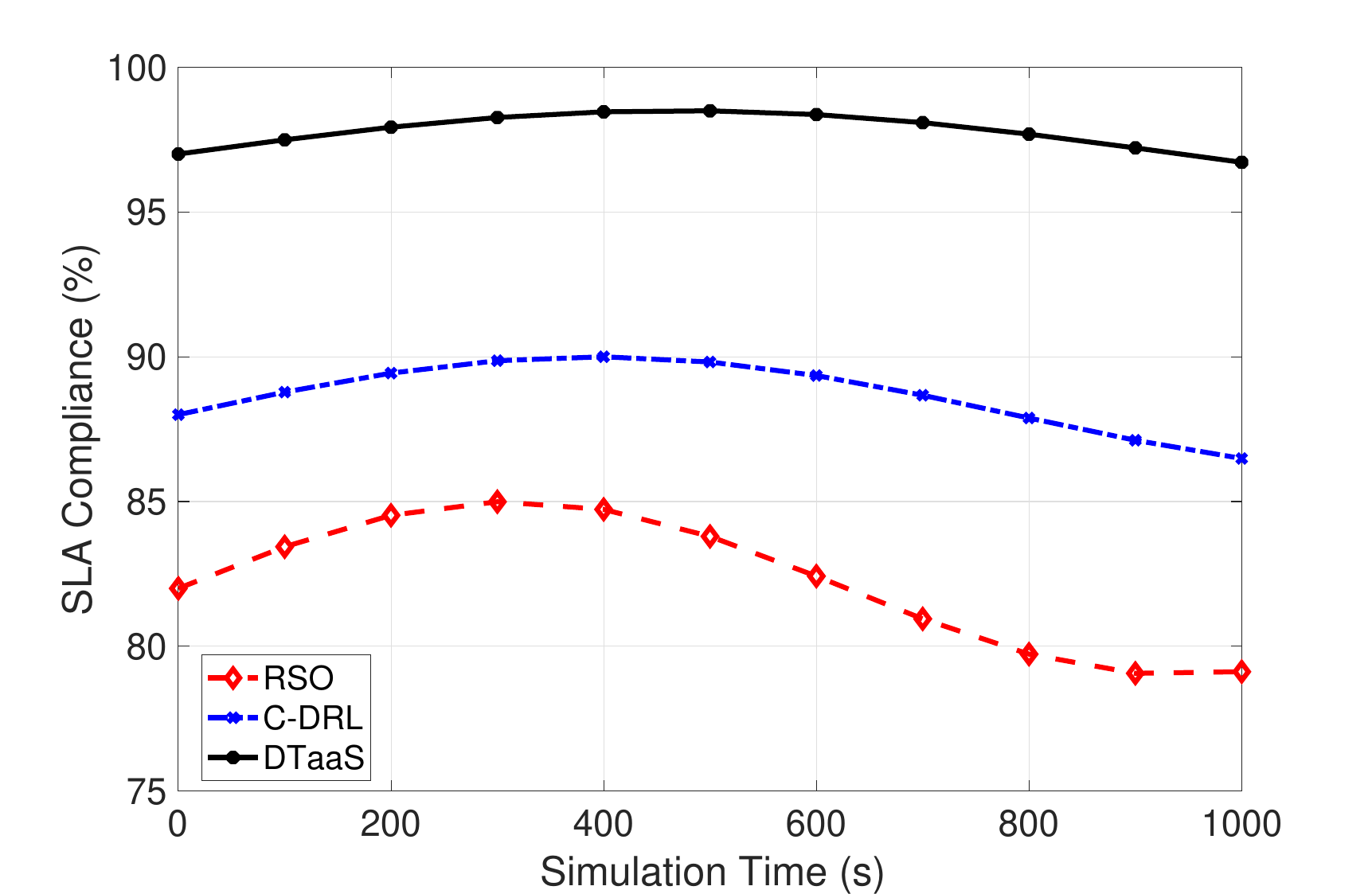}%
		\label{g1}%
	}  %
        \subfloat[]{%
		\includegraphics[width=0.25\textwidth]{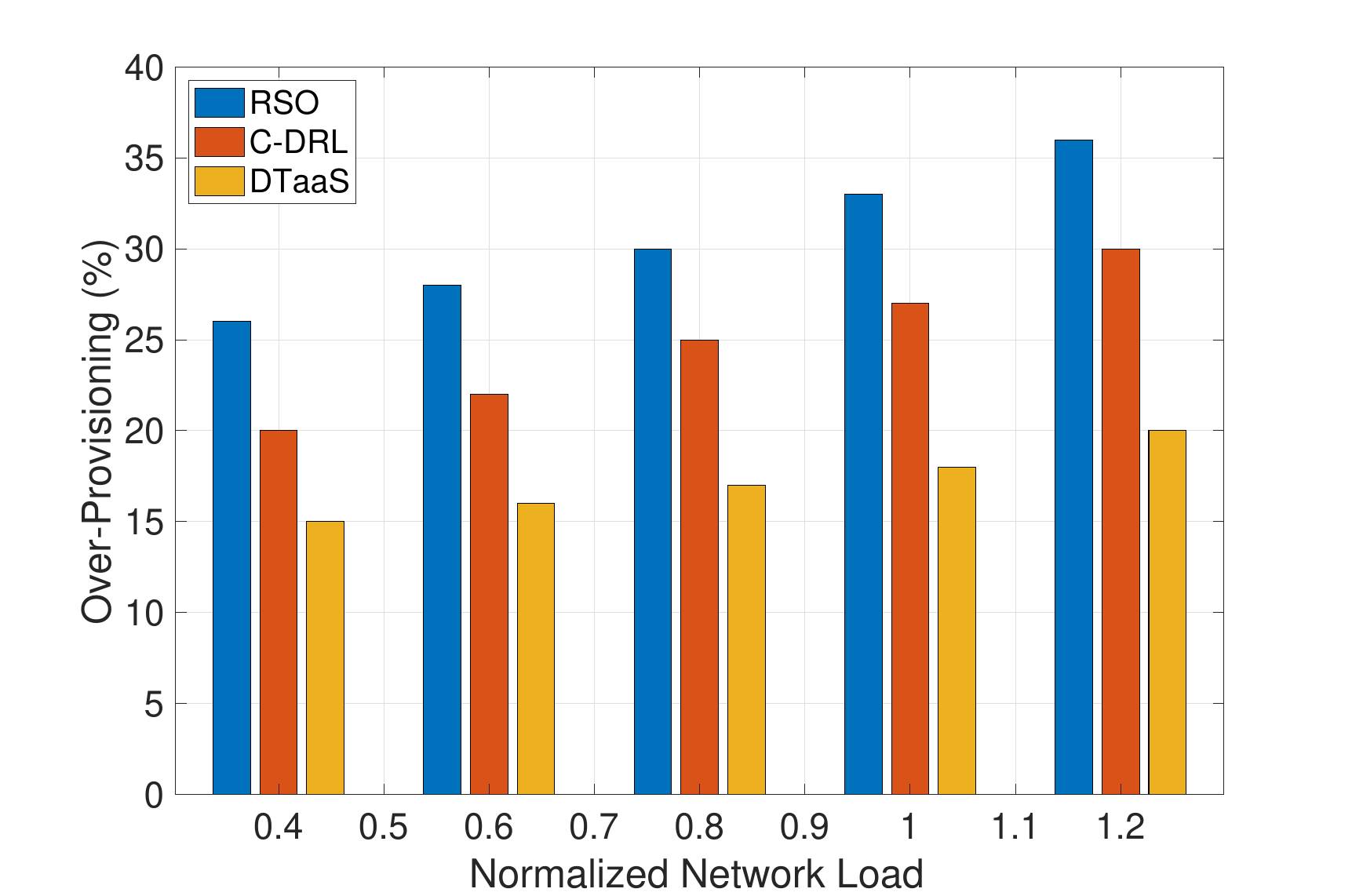}%
		\label{g2}%
	} %
     \subfloat[]{%
		\includegraphics[width=0.25\textwidth]{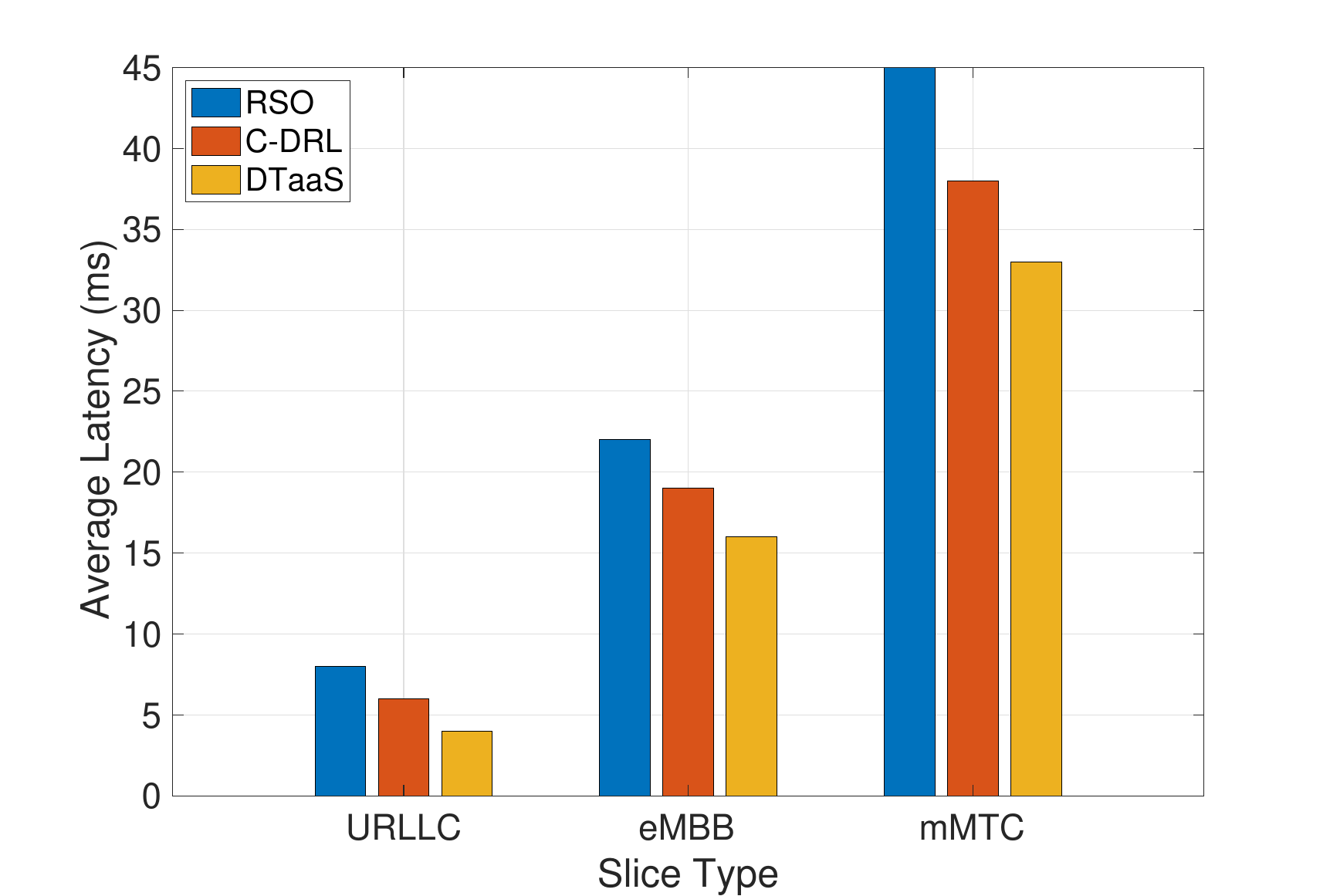}%
		\label{g3}%
	} \\
    \subfloat[]{%
		\includegraphics[width=0.25\textwidth]{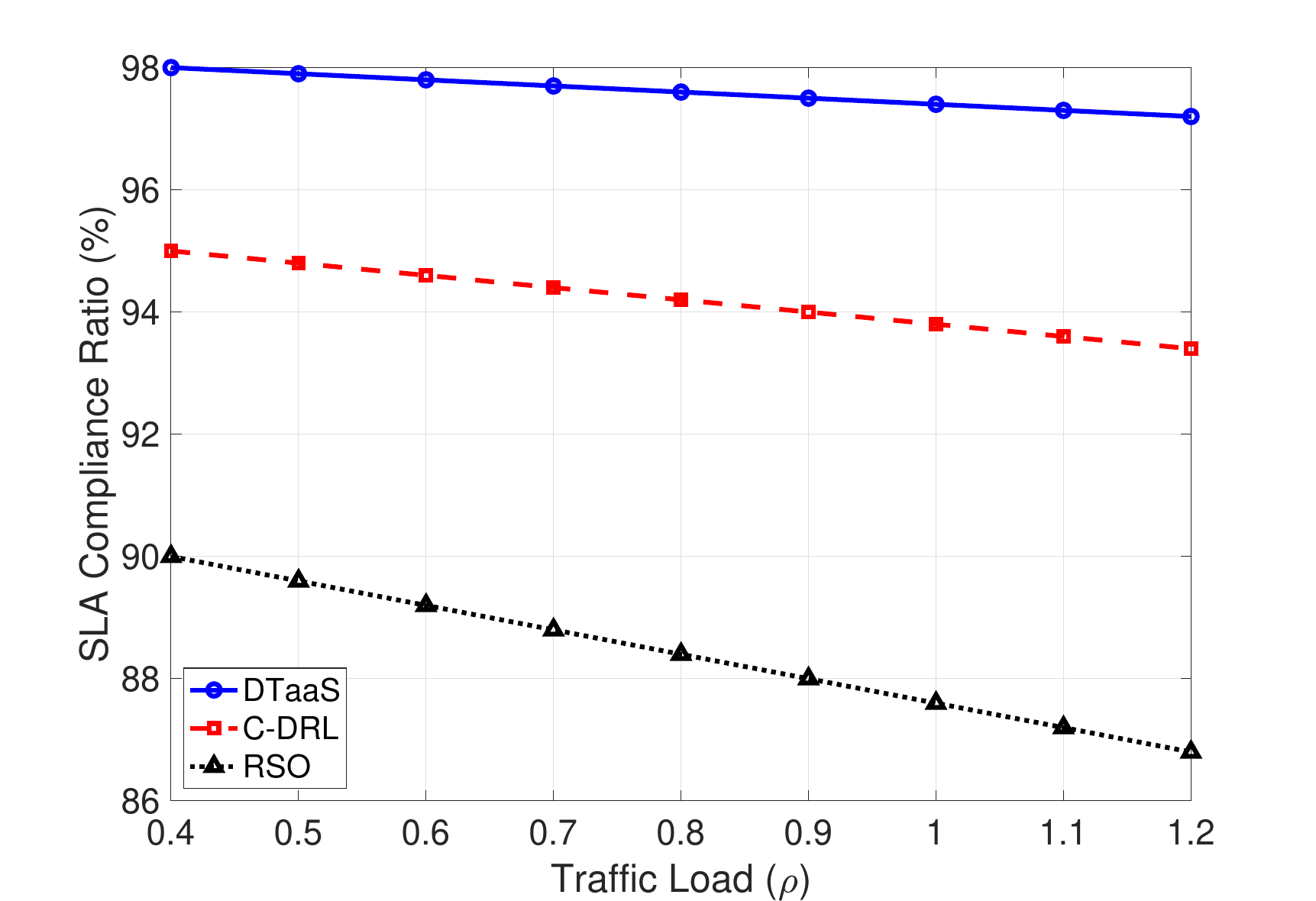}%
		\label{g4}%
	}  %
        \subfloat[]{%
		\includegraphics[width=0.25\textwidth]{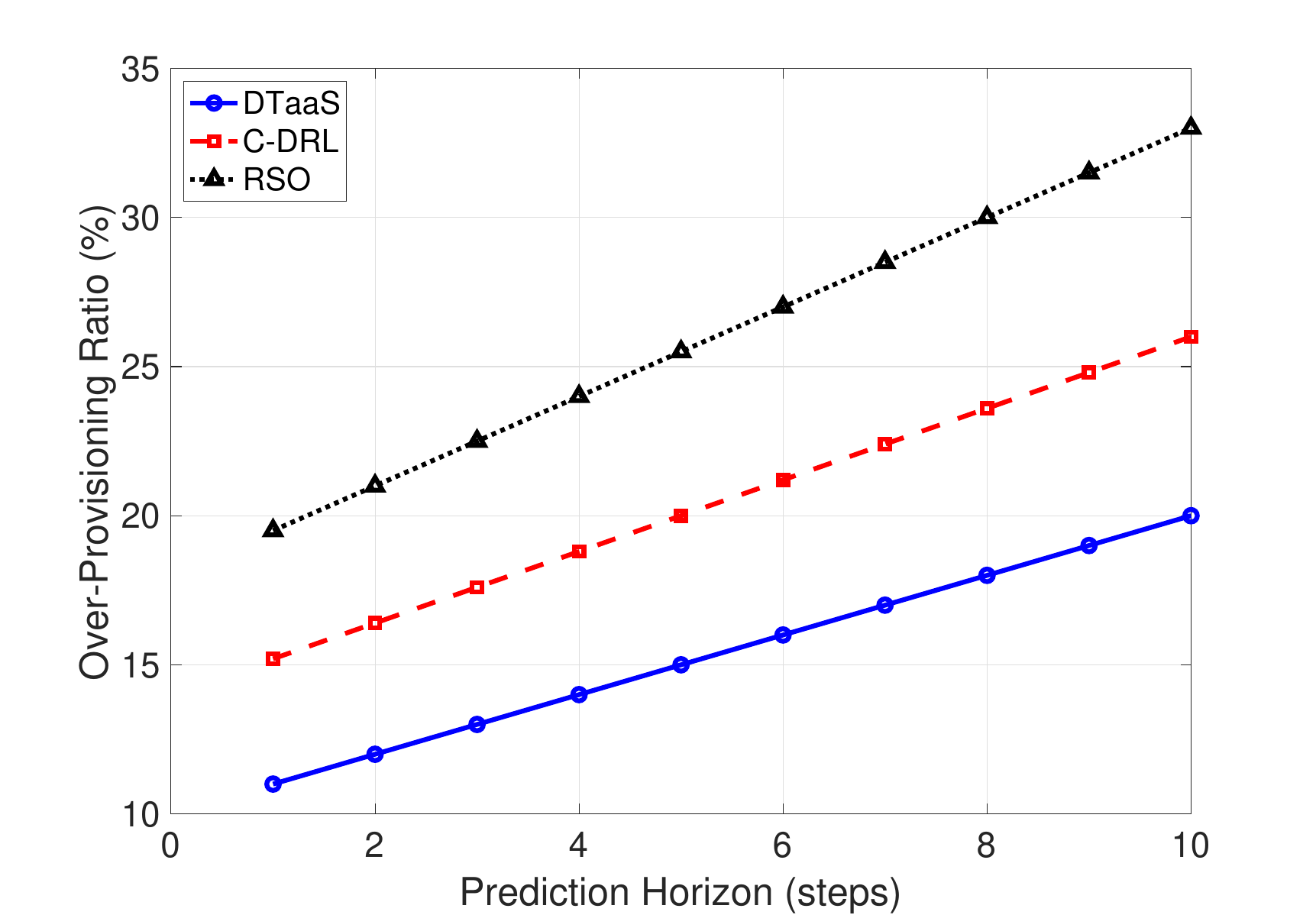}%
		\label{g5}%
	} %
     \subfloat[]{%
		\includegraphics[width=0.25\textwidth]{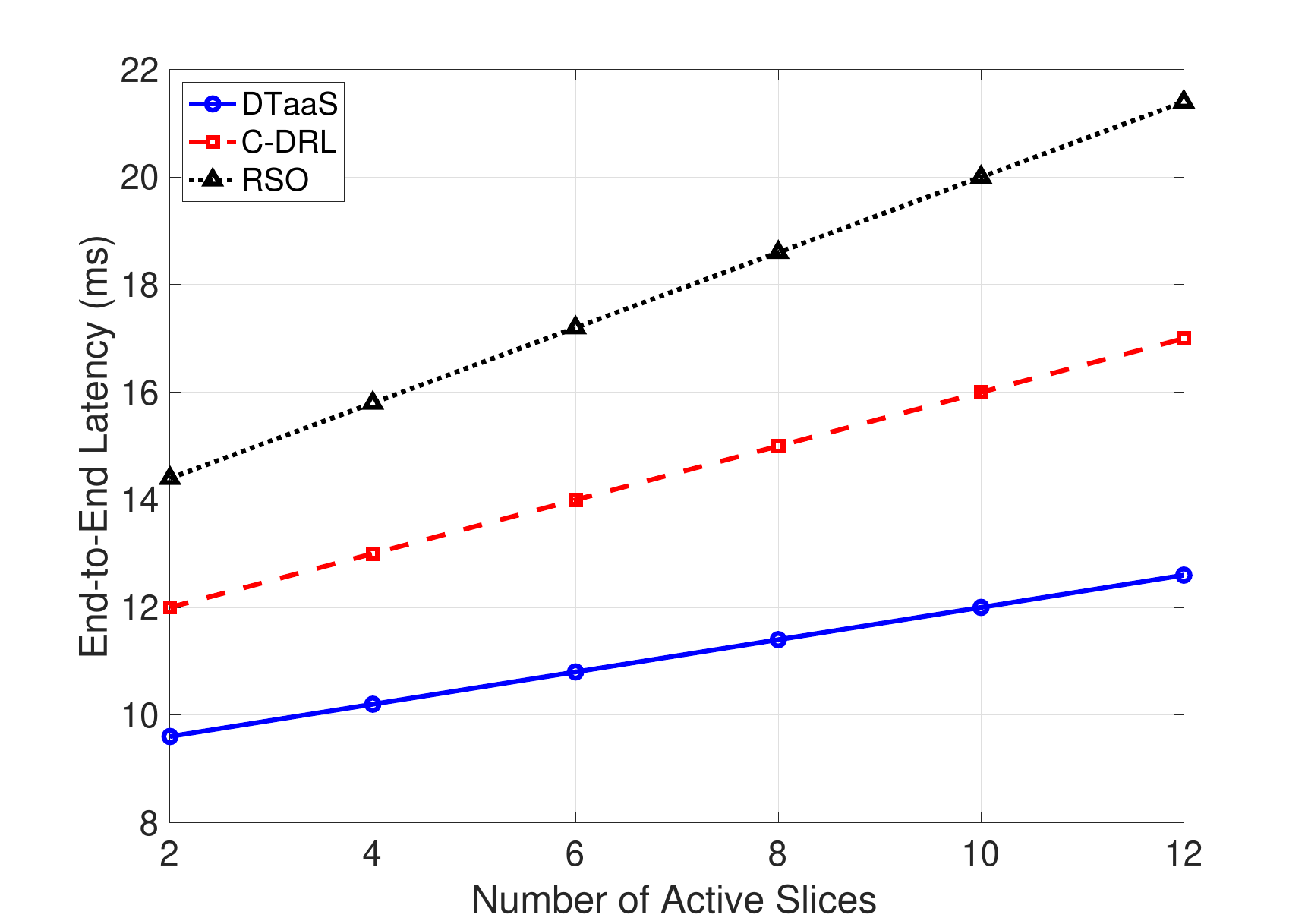}%
		\label{g6}%
	} %
	\caption{Evaluation results.}
	\label{g12}
\end{figure*}

\subsubsection{Evaluation Metrics and Parameters}
The evaluation focuses on three performance metrics:  
\begin{itemize}
    \item {SLA Compliance Ratio (\%)}: It is the percentage of time slots during which slice performance remains above its SLA threshold.
    \item {Resource Over-Provisioning (\%)}: It is the ratio of allocated resources exceeding the minimum required level.
    \item {Average Latency (ms)}: It is the end-to-end service delay, measured from packet arrival to completion.
\end{itemize}

\subsection{Simulation Reslts}

\subsubsection{SLA Compliance Ratio over Time}
As illustrated in Fig.~\ref{g1}, the proposed DTaaS framework consistently maintains a higher SLA compliance ratio compared to both baselines. While the RSO exhibits frequent oscillations due to delayed decision-making, and the C-DRL agent achieves moderate stability with occasional violations under dynamic load, DTaaS sustains over 96\% SLA satisfaction across all simulation periods. This improvement stems from its predictive twin models that forecast short-term traffic fluctuations and preemptively adjust resources, thereby ensuring proactive slice maintenance even during rapid state transitions. Also, Fig.~\ref{g4} illustrates the variation of SLA compliance ratio with normalized traffic load $\rho$. As $\rho$ increases from 0.4 to 1.2, all methods experience a gradual performance drop. However, DTaaS consistently maintains over 95\% compliance due to its predictive control and slice-specific digital twins, which anticipate demand changes and proactively adjust resource allocations. In contrast, the RSO shows a steep decline beyond $\rho=1.0$, while the C-DRL model remains moderately stable but less adaptive.

\subsubsection{Over-Provisioning vs. Network Load}
Figure~\ref{g2} presents the over-provisioning ratio as a function of normalized traffic load. As load intensity increases, both RSO and C-DRL baselines exhibit sharp rises in redundant allocation to preserve SLA compliance, reaching nearly 40\% excess provisioning under peak demand. In contrast, DTaaS significantly reduces over-provisioning through its demand-aware prediction and closed-loop control. The framework maintains optimal utilization, lowering resource wastage by approximately 35–45\% compared to reactive methods. These results confirm that predictive orchestration can simultaneously sustain performance and minimize inefficiency. Moreover, as shown in Fig.~\ref{g5}, the over-provisioning ratio increases with the prediction horizon, reflecting the growing uncertainty of long-term forecasting. DTaaS effectively mitigates this rise by employing sequential models with adaptive horizon control, resulting in up to 30\% less resource redundancy compared to RSO and 18\% less than C-DRL. This demonstrates that DTaaS achieves a better balance between prediction depth and operational efficiency, ensuring scalability without excessive resource reservation.

\subsubsection{End-to-End Latency under Dynamic Traffic}
As shown in Fig.~\ref{g3}, DTaaS achieves substantially lower end-to-end latency, especially under high-load conditions. The centralized nature of C-DRL introduces decision delays, while the threshold-triggered RSO approach causes bursty reconfigurations that transiently degrade latency. By distributing intelligence to the edge and continuously synchronizing slice digital twins with real-time telemetry, DTaaS minimizes orchestration delay and avoids abrupt state transitions. On average, latency is reduced by 28\% relative to C-DRL and 42\% compared to RSO, validating the scalability and responsiveness of the proposed architecture. Also, Fig.~\ref{g6} shows the average end-to-end latency as the number of concurrent network slices increases. The centralized C-DRL agent experiences a linear latency growth due to computation bottlenecks, and RSO exhibits higher fluctuations from its event-triggered reconfigurations. DTaaS, with distributed orchestration at the edge and real-time synchronization of digital twins, maintains the lowest latency profile, improving responsiveness by 25–40\% over baselines under heavy slicing density.

Overall, the experimental results demonstrate that the DTaaS framework outperforms traditional reactive and centralized learning-based approaches across all key performance indicators. By integrating multi-domain telemetry, predictive modeling, and distributed orchestration, DTaaS achieves a balanced trade-off between SLA reliability and resource efficiency. The results confirm that embedding per-SDTs into the orchestration loop enables closed-loop, proactive control suitable for dynamic 6G slicing environments.

\section{Conclusion}
This paper presented a DTaaS framework for predictive slice management in 6G networks. The proposed architecture embeds per-SDTs as active orchestration entities, integrating multi-domain telemetry, sequential predictive models, and distributed edge intelligence. Through simulation-based evaluations, DTaaS demonstrated superior SLA compliance, lower latency, and reduced over-provisioning compared to reactive and centralized learning-based baselines. These results validate the feasibility of proactive, closed-loop orchestration for highly dynamic and heterogeneous 6G environments.

%In future work, we plan to extend the DTaaS framework toward real-world testbed deployment with heterogeneous access technologies, such as terrestrial and non-terrestrial (NTN) segments. Integration with intent-based networking (IBN) and federated learning will also be explored to enhance scalability and privacy in distributed orchestration. Additionally, incorporating energy efficiency and sustainability metrics into the DT-driven decision process will be a key direction to align DTaaS with the green-6G vision. By advancing these dimensions, we aim to transition DTaaS from simulation to a deployable, AI-native orchestration paradigm for next-generation intelligent networks.

\bibliographystyle{IEEEtran}
\bibliography{ref2}

\end{document}